\documentclass[10pt]{article}

\usepackage{amsmath}
\usepackage{amssymb}

\usepackage{graphicx}

\usepackage{cite}

\usepackage{color} 

\usepackage[labelfont=bf,labelsep=period,justification=raggedright]{caption}

\makeatletter
\renewcommand{\@biblabel}[1]{\quad#1.}
\makeatother

\date{}

\begin{document}

\begin{flushleft}
{\Large
\textbf{Early warning signals for critical transitions: A generalized modeling approach}
}
\\
Steven J. Lade$^{\ast}$, 
Thilo Gross
\\
Max Planck Institute for the Physics of Complex Systems, N\"othnitzer Str. 38, 01187 Dresden, Germany
\\
$\ast$ E-mail: slade@pks.mpg.de
\end{flushleft}

\section*{Abstract}
Critical transitions are sudden, often irreversible, changes that can occur in a large variety of complex systems. 
Signals that warn of critical transitions are highly desirable, but their construction can be impeded by limited availability of data. 
We propose a method that can significantly reduce the amount of time series data required for a robust early warning signal by using other information about the system. 
This information is integrated through the framework of a generalized model. 
We demonstrate the applicability of the proposed approach through several examples, including a previously published fisheries model.

\section*{Author Summary}
Fisheries, coral reefs, productive farmland, planetary climate, neural activity in the brain, and financial markets are all complex systems that can be susceptible to sudden changes leading to drastic re-organization or collapse.
A variety of signals based on analysis of time-series data have been proposed that would provide warning of these so-called critical transitions. The amount of data these approaches require, however, can be prohibitive, especially in ecological contexts.
We propose a new method for calculating early warning signals that can significantly reduce the amount of data required. The key step is to incorporate other readily available information about the system through the framework of a so-called generalized model.
Our new approach may help to anticipate future catastrophic regime shifts in nature and society, allowing humankind to avert or to mitigate the consequences of the impending change.

\section*{Introduction}
Critical transitions are sudden, long-term changes in complex systems that occur when a threshold is crossed \cite{Scheffer_2009}.
Many systems are known to be at risk of such transitions, including systems in ecology \cite{Scheffer_N_2001}, climate research \cite{Lenton_PNAS_2008}, economics \cite{May_N_2008}, sociology \cite{Brock_2006} and human physiology \cite{Venegas_N_2005}. 
Examples of critical transitions in ecology include shifts in food web composition in shallow lakes \cite{Carpenter_S_2011}, degradation of coral reefs \cite{Mumby_N_2007}, degradation of managed rangelands \cite{Anderies_E_2002}, and desertification \cite{Foley_E_2003}. 

Warning signals for impending critical transitions are highly desirable, because it is often difficult to revert a system to the previous state once a critical transition has occurred \cite{Scheffer_N_2001,Folke_AREES_2004}.
If an accurate mathematical model of the system is available then critical transitions can be predicted straight-forwardly, either by numerical simulation or by direct computation of the dynamical thresholds. 
For real world complex systems, however, sufficiently accurate models are in general not available, and predictions based on models of limited accuracy face substantial difficulties \cite{Groffman_E_2006}. 
Recent research has therefore focused on model-free approaches that extract warning signals from observed time series \cite{Scheffer_N_2009}.
Two of the most widely used approaches are increasing variance \cite{Carpenter_EL_2006} and autocorrelation \cite{Dakos_PNAS_2008}, both of which are caused by critical slowing down \cite{Wissel_O_1984}. 
Other approaches consider warning signals based on skewness \cite{Guttal_EL_2008}, flickering \cite{Brock_EM_2010} and spatial correlation \cite{Dakos_TE_2010}.

In practice, robust calculation of early warning signals such as these is often limited by the need to acquire sufficient data. 
Especially in ecological applications, acquiring time series with the desired accuracy and sampling rate poses a significant challenge. 
One strategy for reducing the demand for time series is to utilize other information that may be available. 
This highlights the need for intermediate approaches, which can efficiently incorporate available information without requiring a fully specified mathematical model.   

In the present Letter, we propose an approach for the prediction of critical transitions based on the framework of generalized modeling \cite{Gross_PRE_2006,Gross_Science_2009}. 
The approach allows available information to be used to reduce the amount of time series data required or to increase the quality of the prediction.  
We demonstrate the applicability of the proposed approach by considering a simple one-population model, a previously studied fisheries model and a tri-trophic food chain.

\section*{Methods}
\subsection*{Generalized modeling}
Suppose that a system has been identified as being at risk of a critical transition. 
Even if very little specific information is available, the dynamics can generally still be captured by a so-called generalized model \cite{Gross_PRE_2006}. 
Such a model captures the structure of the system, without restricting it to specific functional forms. 

To formulate a generalized model we first identify important system variables (say, abundance or biomass of the populations in the system) and processes (for example, birth, death, or predation). 
As a first step, the generalized model can then be sketched in graphical form, such as in Fig. \ref{fig:Biggsschematic} below. 
This graphical representation is sometimes called a causal loop diagram \cite{Sterman_2000}.

To obtain a mathematical representation of the model we write a dynamical equation for each variable $X_i$. 
In these equations we represent the processes by general functions. 
For instance we can model a single population $X_1$ subject to gains $G$ and losses $L$ by an ordinary differential equation
\begin{equation*}
\frac{\rm d}{\rm dt} X_1 = G(X_1) - L(X_1),  
\end{equation*} 
or as a discrete-time map
\begin{equation*}
X_{1,t+1} = G(X_{1,t}) - L(X_{1,t}).  
\end{equation*} 
Note, that in contrast to conventional models, we do not attempt to describe the processes G and L by specific functional forms. 
Instead, we use unspecified functions $G()$ and $L()$ as formal placeholders for the (unknown) relationships realized 
in the real system.   

\subsection*{Calculation of early warning signal}
We assume that before the critical transition, the system can be considered in equilibrium. 
We emphasize that this does not require the system to be completely static or closed in a thermodynamic sense, 
but that, on the chosen macroscopic level of description, the system can be considered at rest.
Additionally, the system is subject to a slowly changing external parameter 
that puts it at risk of undergoing a critical transition. 
The system is therefore at equilibrium only on a certain timescale. 
In the following we refer to this timescale as the \emph{fast timescale}, while the dynamics of the 
whole system, including the slow change of the external parameter, takes place on the \emph{slow timescale}.  

Using the definitions above critical transitions can be linked to instabilities (bifurcations) of the fast subsystem \cite{Kuehn_2010}. 
For detecting these instabilities we construct the Jacobian matrix, a local linearization of the system around the steady state \cite{Kuznetsov_2010}.
A system of ordinary differential equations (ODEs) is dynamically stable if all eigenvalues of the Jacobian have negative real parts, 
whereas a discrete time map is stable if all eigenvalues have an absolute value less than one.  
Critical transitions are thus signified by a change in the external parameter causing at least one of the eigenvalues to 
cross the imaginary axis (ODE) or a unit circle around the origin (map).

To warn of impending critical transitions we monitor the eigenvalues of the Jacobian of the fast subsystem, which usually 
change slowly in time. A warning is raised if at least one of the eigenvalues shows a clear trend toward the stability boundary 
(for ODEs, zero real part; for maps, absolute value of one). 
The Jacobian itself can be computed directly from the generalized model, but will contain unknown terms reflecting our ignorance of the precise functional forms in the model. 
Previous publications \cite{Gross_PRE_2006} have shown that these unknowns can be treated as well-defined parameters with clear 
ecological interpretations. 
In the present applications we estimate the unknowns in the Jacobian matrix from short segments of time series data.
Thereby, a pseudo-continuous monitoring of the eigenvalues of the fast subsystem is possible.  

We note that with given time series data estimating the generalized model parameters is simpler than estimating the entries of the Jacobian matrix directly, because the generalized model already incorporates structural information about the system. Further, 
many of the parameters in the generalized model may already be available in a given application, because they refer to well-studied properties of the species, such as natural life expectancy or metabolic rate. 

\section*{Results}
We applied the proposed approach to three case studies, focusing on 
a generic population with an Allee effect, a fisheries example, and a tri-trophic food chain.  

\subsection*{Simulation with Allee effect}
Allee effects, that is, positive relationships between per-capita growth rate and population size, are postulated in many populations and have been conclusively demonstrated in some \cite{Kramer_PE_2009}. 
A population with an Allee effect can suddenly transition from a stable, non-zero population size to unconditional extinction \cite{Boukal_JTB_2002}. 
We constructed the Jacobian of such a system from a suitable generalized model (see Appendix S1 in Supporting Information).
This Jacobian contains unknown parameters that depend on the specific system under consideration. 
An early warning signal is obtained by estimating these parameters from time series data of the population size and birth rate, and 
monitoring the eigenvalues of the Jacobian as external parameters change.  

For testing the early warning signal, described above, we simulated a simple model
\begin{equation}
\label{eq:Allee_actual}
\frac{\rm d}{\rm dt}X = \frac{A X^2}{k^2 + X^2} - mX + \sigma \xi(t),
\end{equation}
based on Yeakel \emph{et al.} \cite{Yeakel_TE_2011}. We included a small additive noise term $\sigma \xi(t)$, with standard deviation $\sigma$ and autocorrelation $\langle \xi(t) \xi(t') \rangle = \delta(t-t')$.
Throughout the simulation we slowly changed the mortality rate $m$ according to $m = 7.5 + 0.2t$. A critical transition occurred, causing subsequent extinction of the population (Fig.~\ref{fig:Allee}). 
 
The challenge we address is predicting the critical transition from a limited number (here, fifteen) of observations of population size and birth rate. 
We emphasize that we do not utilize any information on the functional forms of processes employed in the simulation, so
that the prediction is based solely on the 15 observations and the assumed structural information (that is, one population subject to gains and losses).
By estimating the parameters of the generalized model we determined the eigenvalues of the Jacobian as a function of time (Fig. \ref{fig:Allee}b).
A clear increase in the eigenvalue is detectable well before the critical transition, giving ample warning of the impending collapse.

Due to a phenomenon called bifurcation delay \cite{Kuehn_2010}, the population size did not start to change rapidly until well after ($t \approx 13.5$) the bifurcation point ($t = 12.5$). As previously observed by Biggs \emph{et al.} \cite{Biggs_PNAS_2009}, management action to reverse the change in bifurcation parameter may successfully avert the critical transition even after the fast subsystem's bifurcation has occurred, if still within the range of the bifurcation delay. In the case of Fig. \ref{fig:Allee}(b), the eigenvalue trend is directed more towards the last possible time that successful management action can be taken than towards the time of the actual bifurcation.

\subsection*{Fishery simulation of Biggs \emph{et al.}}
Our second case study focuses on an example from fisheries. 
Increased harvesting of piscivores can induce a shift from the high-piscivore low-planktivore regime to a low-piscivore high-planktivore regime \cite{Walters_CJFAS_2001}. 
Many fisheries are suspected to have undergone such transitions \cite{Steele_PO_2004,deYoung_TEE_2008}. 
Based on the causal loop diagram (Fig.~\ref{fig:Biggsschematic}), we formulated a discrete-time generalized model, describing 
the piscivore and planktivore populations at the end of each year, in the spirit of stock-assessment modeling (see Appendix S1). 
Thereby detailed modeling of the intra-annual dynamics was avoided. 
 
To test the warning signal, we generated time series data with a detailed fishery model by Biggs \emph{et al.} \cite{Biggs_PNAS_2009}, which was developed from a series of whole-lake experiments \cite{Carpenter_EL_2008}. 
We note that this model differs significantly from our generalized model by a) accounting for the intra-annual dynamics and b) 
containing an additional state variable denoting the juvenile piscivore population. 
These discrepancies were intentionally included to reflect the limited information that would be available for the formulation of the generalized model in practice.
In simulations the detailed model showed a transition to a low-piscivore high-planktivore regime as the harvesting rate was increased (Fig. \ref{fig:Biggs}a). 

From this simulation, we recorded the simulated piscivore and planktivore density and piscivore catch at the end of each year.
Because the simulated data was very noisy we estimated the Jacobian's eigenvalues after smoothing the recorded data (see Appendix S1).
In addition to the time series data, the information on the natural adult piscivore mortality and reproduction rate and the planktivore influx from refugia were required (see Appendix S1). This type of information can be reasonably well estimated for most systems. 
We confirmed that our predictions (reported below) are not sensitive to the specific values used. 
Indeed, a simple approach for estimating these parameters is to recognize that the initial state, before the critical transition, is stable. 
In a number of test trials we confirmed that any reasonable combination of parameters used that corresponded to an initially stable steady state provided an early warning signal comparable to the results reported below.

An estimate of the Jacobian eigenvalues for the fisheries example is shown in Fig.~\ref{fig:Biggs}b. 
As the system approaches the critical transition we observe that an eigenvalue approaches one, which signifies a critical transition for discrete time systems. 
This result is compared to the variance-based early warning signal computed by Biggs \emph{et al.} \cite{Biggs_PNAS_2009}, which uses a much more densely sampled time series including intra-annual dynamics.
The comparison shows that the approach proposed here produces a signal of similar quality (although possibly \emph{too} early), while requiring significantly less time series data. 
Further, comparison with a variance-based signal using the same amount of time-series data as the generalized model shows that the generalized model-based signal is a much clearer early warning signal in this case. In particular, the variance signal only rises during or after the transition.

\subsection*{Tri-trophic food chain}
For our final example we consider a tri-trophic food chain.
In ecological theory food chains play a role both as a prominent motif appearing in complex food webs and as coarse-grained models. 
Using generalized models, a general Jacobian for a continuous-time model of the tri-trophic food chain can be derived (see Appendix S1 and Gross \emph{et al.} \cite{Gross_O_2005}).   

For generating example time series data we used a conventional equation-based model in which a producer biomass, $X_1$,
predator biomass, $X_2$, and top predator biomass, $X_3$, follow  
\begin{subequations} \label{eq:tri_actual}
\begin{align}
\frac{\rm d}{\rm dt} X_1&= A_n X_1 (K_n - X_1) - \frac{B X_1^2 X_2}{K_3 + X_1^2} + \sigma_1 \xi_1(t) \\
\frac{\rm d}{\rm dt} X_2&= \frac{B X_1^2 X_2}{K_3 + X_1^2} - \frac{A_p X_2^2 X_3}{K_p+X_2^2} + \sigma_2 \xi_2(t) \\
\frac{\rm d}{\rm dt} X_3&= \frac{A_p X_2^2 X_3}{K_p+X_2^2} - mX_3 + \sigma_3 \xi_3(t).
\end{align}
\end{subequations}
In these equations we assumed Holling type-3 predator-prey interaction, logistic growth of the producer and linear mortality of the top predator and $\xi_i(t)$ are additive noise terms with $\langle \xi_i(t) \xi_j(t') \rangle = \delta_{ij} \delta(t-t')$. 
If any biomass decreased to zero, we suppressed the noise term so that the corresponding population remained extinct.

We simulated the Eqs.~\eqref{eq:tri_actual} while increasing the mortality rate of the top predator. 
The resulting time series, for the chosen combination of parameters, show a slowly changing steady-state followed by a sudden transition to large oscillations, and a
sudden collapse of all three populations (Fig.~\ref{fig:tri}). 

To provide an early warning signal for the transition we recorded time series of the three biomasses and the top-predator's death rate, and estimated the parameters of the generalized model from smoothed segments of these time series. 
Even for the smoothed data we find that one of the eigenvalues is very noisy and sometimes positive. 
We believe that the presence of this eigenvalue reflects the response of the prey to fluctuations on the higher 
trophic levels and therefore exclude this value from our analysis. 
As $m$ is increased toward the critical transition at the onset of oscillations, two eigenvalues show a clear increase 
toward zero real part (Fig.~\ref{fig:tri}). 
The two eigenvalues approach zero as a complex conjugate eigenvalue pair, which is indicative of the system undergoing 
a Hopf bifurcation \cite{Kuznetsov_2010}. 
This bifurcation generally implies a transition from stationary to oscillatory dynamics. The large oscillations combined with stochastic fluctuations then lead rapidly to extinction.
The early warning signal constructed here thus indicates the transition to an oscillatory state significantly before the transition occurs.

\section*{Discussion}
In this Letter we proposed an approach for anticipating critical transitions before they occur. 
In particular we showed that generalized modeling of the system can facilitate the incorporation of the structural information that is in general available.

We demonstrated the proposed approach in a series of three case studies. 
The first example showed that in simple systems even very few time points can be sufficient for clean prediction of the critical transition. 
The second example posed a hard challenge, where test data was generated by a model that differed considerably from the generalized model. 
Yet even in this case the generalized model significantly reduced the amount of data needed for predicting the transition. 
The third and final example demonstrated the ability of the proposed approach to distinguish between different types of critical transitions. 
 
In all case studies we found that the proposed approach can robustly warn of critical transitions in the presence of noise. 
We believe that the performance of the approach under noisy conditions can be further improved by subsequent refinements. Such refinements could include
combination with dynamic linear modeling \cite{Carpenter_EL_2006}, utilization of a parameter transformation (to `scale' and `elasticity' parameters) previously proposed for generalized models \cite{Gross_PRE_2006}, or the use of optimized sampling procedures. 

An advantage of the proposed approach is that it is generally becomes more reliable closer to a critical transition, where rates of change of state variables and other observables are generally larger, leading to better sampling. 
In contrast, methods based on variance or autocorrelation become statistically more difficult to estimate as the time series becomes less stationary. In this sense the proposed approach provides a tool complementary to established variance-based methods.  

An important assumption in our present treatment was that the dynamics of the fast subsystem could, at least at some level of description, be considered as stationary. Let us emphasize that this is not a strong assumption because even systems that 
are primarily non-stationary, such as the fisheries example, can be modeled as stationary if a suitable modeling framework is chosen. 
Furthermore, ongoing efforts aim at extending the framework of generalized modeling to non-stationary dynamics, which may lead to a further relaxation of that assumption in the future \cite{Kuehn_2011}. 

In summary, we used generalized models to efficiently incorporate available information about a system without requiring detailed knowledge about that system. 
This led to the construction of early warning signals for critical transitions that required significantly less time series data than variance-based approaches. Thereby the proposed approach can contribute to the warning of critical transitions from a realistic sampling effort.

\section*{Acknowledgments}
The authors acknowledge useful discussions with Christian Kuehn.

\newpage
\begin{figure}[!htp]
\begin{center}
\includegraphics[width=3in]{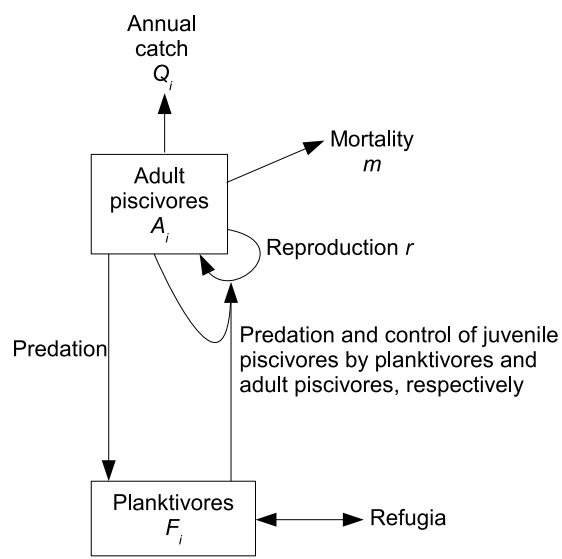}
\end{center}
\caption{
{\bf  Schematic of the fishery knowledge that was incorporated into the generalized model.}  
}
\label{fig:Biggsschematic}
\end{figure}

\newpage
\begin{figure}[!htp]
\begin{center}
\includegraphics[width=4in]{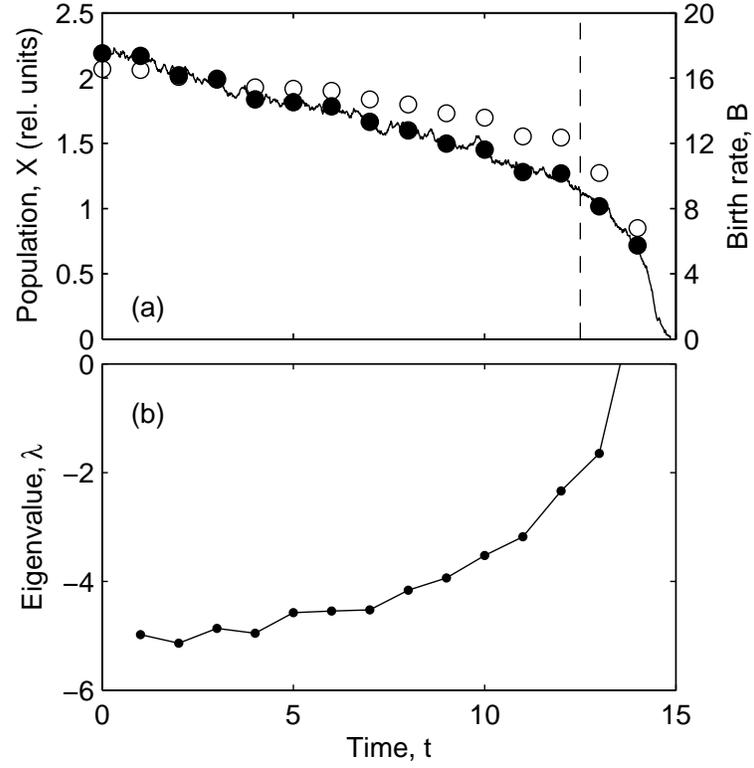}
\end{center}
\caption{
{\bf Early warning signal for a single population with Allee effect.}  (a) Population time series (in relative units) generated by Eq.~\eqref{eq:Allee_actual} (solid line, left axis), and yearly births (circles, right axis). The vertical dashed line indicates the time of the bifurcation. Parameters were $A = 20$, $k = 1$, $m = 7.5 + 0.2t$ and $\sigma = 0.1$. The simulation was started at $t < 0$ to allow for the decay of transient responses. (b) Eigenvalues estimated from the sampled data indicated with markers in (a), using the procedure described in the text. The eigenvalue was always real.  
}
\label{fig:Allee}
\end{figure}

\newpage
\begin{figure}[!htp]
\begin{center}
\includegraphics[width=4in]{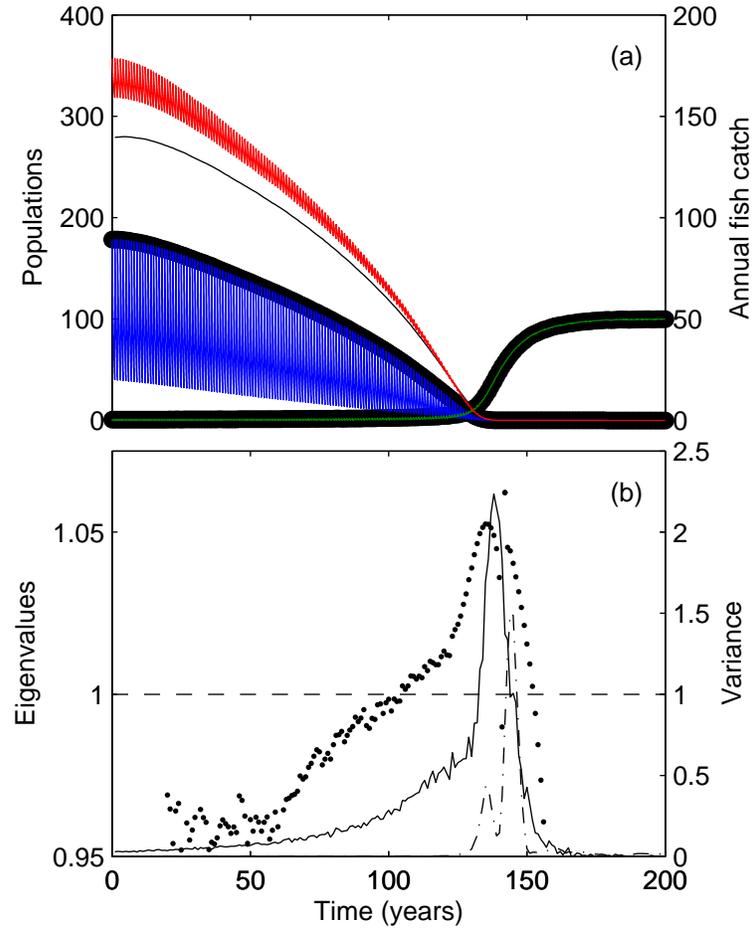}
\end{center}
\caption{
{\bf Early warning signals for the fishery simulation of Biggs \emph{et al.} \cite{Biggs_PNAS_2009}.}  (a) Results for the adult piscivore (blue line, left axis), juvenile piscivore (red line, left axis) and planktivore (green line, left axis) populations of the model of Biggs \emph{et al.} \cite{Biggs_PNAS_2009}, with the harvesting rate they specified of $1.5 + 0.005t$ per year. From these results, data were sampled from $A$ and $F$, once per year, after maturation and mortality for that year had been computed (black markers/thick black line). The annual piscivore catch data used in the early warning signal calculation are also shown (thin black line, right axis). (b) Early warning signals: the eigenvalues (points, left axis) estimated by the generalized modeling approach; the intra-annual variance of the planktivore population (solid line, right axis) as calculated by the method of Biggs et al. \cite{Biggs_PNAS_2009}; and the planktivore variance using only year-end data (dot-dashed line, right axis). Only one of the eigenvalues is visible at this scale. The eigenvalues were always real. The year-end variance warning signal was calculated using a sliding window of the previous 15 year-end planktivore populations and a quadratic detrending within that window, which yielded generally stationary data for those windows that were before the critical transition.  
}
\label{fig:Biggs}
\end{figure}

\newpage
\begin{figure}[!htp]
\begin{center}
\includegraphics[width=3.4in]{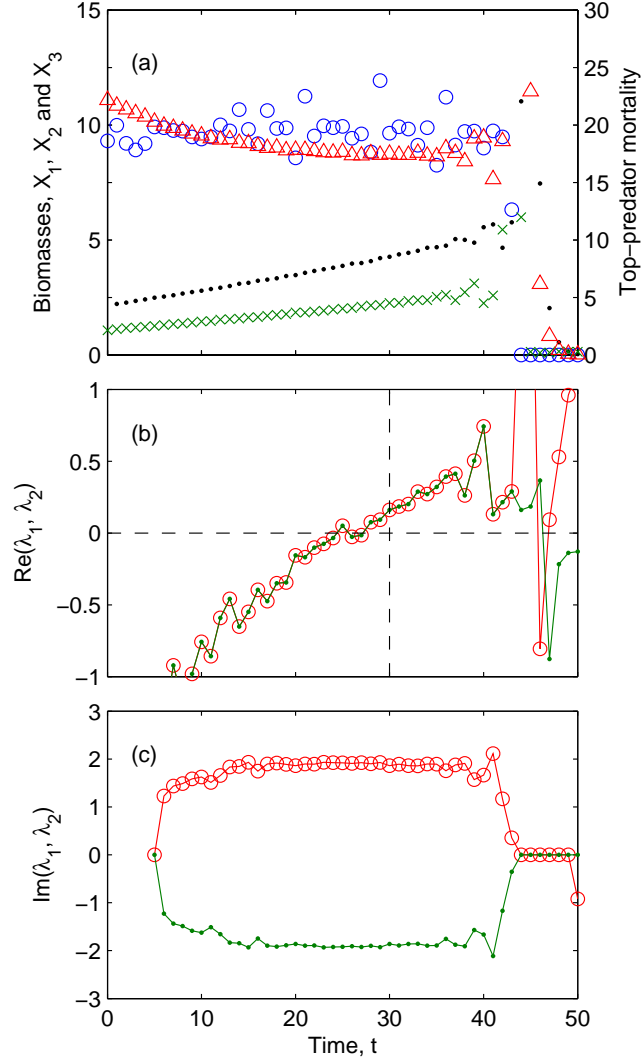}
\end{center}
\caption{
{\bf Early warning signal for a critical transition in a tri-trophic food chain.}  (a) Time series of $X_1$ (blue circles, left axis), $X_2$ (green crosses, left axis) and $X_3$ (red triangles, left axis) generated by Eq.~\eqref{eq:tri_actual}. Only the data subsequently used in the early warning analysis were plotted. Some of the data during the oscillations between $t = 40$ and $45$ were outside the scale of this graph, with $X_3$ exceeding 30. The estimates of top-predator mortality used to calculate the early warning signal are also shown (black dots, right axis). Parameters were $A_n = 4$, $B = 4$, $A_p = 2$, $K_p = 5$, $K_n = 10$, $K_3 = 2$, $(\sigma_1, \sigma_2, \sigma_3) = (5, 0.02, 0.01)$ and $m(t) = 0.4 + 0.02t$. (b,c) Real and imaginary parts of eigenvalues estimated from the data in (a), using the procedure described in the text.
Eigenvalues are denoted by dots and circles; a dot within a circle indicates two eigenvalues had the same real value. The markers and colors used in (b,c) have no correspondence to those used in (a). A third (purely real) eigenvalue was not plotted, for reasons described in the main text. In (b), the horizontal dashed line indicates the stability boundary at zero real part, while the vertical dashed line indicates the time that the (Hopf) bifurcation occurred in the fast subsystem of Eqs.~\eqref{eq:tri_actual}.  
}
\label{fig:tri}
\end{figure}

\newpage
\newpage
\newpage

\begin{flushleft}
{\Large
\textbf{Appendix S1: Detailed method for early warning signal calculation}
}
\\
Steven J. Lade$^{\ast}$, 
Thilo Gross
\\
Max Planck Institute for the Physics of Complex Systems, N\"othnitzer Str. 38, 01187 Dresden, Germany
\\
$\ast$ E-mail: slade@pks.mpg.de
\end{flushleft}

This Appendix details the generalized models, the assumptions behind the generalized models, and the numerical scheme used to generate the early warning signals for the three case studies in the main text.

\section*{Single population with Allee effect}
In generating an early warning signal for critical transition in the population with Allee effect, we assumed:
\begin{itemize}
\item That the equation
\begin{equation}
\label{eq:Allee_X}
\frac{\rm d}{\rm dt}X = B(X) - M(X,\mu)
\end{equation}
is a good generalized model for the system, where $B(X)$ and $M(X,\mu)$ are the birth and death rates of the population, respectively. Of these two processes, only the death rate is affected by the external parameter $\mu$ pushing the system towards the critical transition. This factor could be anything that changes the death rate: the spread of a new disease, the appearance of a new predator, or habitat destruction, for example.
\item Access to a time series of the population $X(t)$. We refer to the population observations as $X_i$, taken at times $t_i$, $i = 1, \ldots, n$.
\item Access to a time series of the birth rates $B_i$ (the death rate $M_i$ would also be acceptable). For simplicity, we assume these observations are also at times $t_i$.
\item That the mortality $M(X,\mu)$ is linear in $X$ (although the coefficient of this linearity may change with $\mu$).
\end{itemize}

From Eq. \eqref{eq:Allee_X}, the eigenvalue of the system's dynamics near its steady state is given by
\begin{equation}
\label{eq:Alleelambda}
\lambda = B'(X) - M'(X,\mu),
\end{equation}
where the prime denotes the derivative with respect to $X$. This equation shows that the gradients $B'$ and $M'$ are sufficient to calculate $\lambda$. Since the birth rate $B_i$ and the population $X_i$ have been directly observed, $B'_i = \Delta B_i/\Delta X_i$ could therefore be computed immediately, where we use the notation $\Delta G_i \equiv G_i - G_{i-1}$. A discretization of Eq.~\eqref{eq:Allee_X} gives $M_i = B_i - \Delta X_i/\Delta t_i$. Because $M$ depends on $\mu$ as well as $X$ its derivative could not be estimated in the same way as $B'$. Using the assumption of linearity of $M$, we estimated $M'_i = M_i/X_i$. (Suppose $M = k X$. Then $M' = k = M/X$.) We could then estimate the eigenvalue $\lambda_i = B'_i - M'_i$.

These derivative estimators are one-sided, in that they approximate $f'(x) \approx \left[f(x) - f(x-\Delta x) \right]/\Delta x$. This one-sided estimation involved a loss in accuracy but allows the eigenvalues to be estimated at the most recent observation time, which is important when attempting to predict an imminent transition.

\newpage
\section*{Fishery simulation of Biggs et al.}
We assumed the following knowledge:
\begin{itemize}
\item The populations of adult piscivores $A_i$ and planktivores $F_i$ at the start of year $i$
\item The number of adult piscivores harvested $Q_i = Q(A_i)$ during year $i$
\item That adult piscivores die from other causes at a per-capita rate $m$
\item That adult piscivores predate on the planktivores
\item That there is a net movement of planktivores into (out of) the foraging arenas from (to) refugia \cite{Walters_2004} of $R_i$ in year $i$
\item That per-capita, and in the absence of interactions with other modeled populations, $r$ piscivores would be born and mature into adult piscivores the following year
\item That however juvenile piscivores are predated on by planktivores \cite{Walters_CJFAS_2001} and are also controlled, for example by accidental predation, by the adult piscivores \cite{Carpenter_EL_2008}.
\end{itemize}
This knowledge is represented schematically in Fig.~1 of the main text.

We constructed a generalized model of this fishery by modeling only the year-end piscivore populations $A_i$ and planktivore populations $F_i$ with a discrete map. As a consequence we could avoid explicitly modeling the complications of intra-annual dynamics such as piscivore reproduction and maturation. We wrote:
\begin{subequations} \label{eq:Biggs} \begin{align}
A_{i+1} &= (1-m)(A_i - Q_i) + rA_i - C_i \label{eq:BiggsA} \\
F_{i+1} &= F_i + R_i - D_i. \label{eq:BiggsF}
\end{align}
\end{subequations}

The first term of Eq.~\eqref{eq:BiggsA} gives the number of surviving adult piscivores after harvesting ($Q_i$) and other causes of mortality ($m$). The second term $rA_i$ models reproduction without losses and the third $C_i = C(A_i,F_i)$ gives the predatory or controlling effects of planktivores and adult piscivores on the number of new adult piscivores. Since $C$ depends twice on the adult piscivore population, first through the number of juvenile piscivores born and again through the controlling effect of the adults, we assume $C$ is quadratic in $A_i$.

In Eq.~\eqref{eq:BiggsF}, $R_i$ models the net influx of planktivores into the foraging arenas. Since the planktivore population was initially small, we assumed there was negligible outflow of planktivores, and that the inflow depended on a refuge size that was constant. Therefore $R_i = R$ was constant. [In the other scenario considered by Biggs \emph{et al.} \cite{Biggs_PNAS_2009}, where habitat destruction causes the refuge size to decrease, $R_i$ would not be constant.] The remaining term $D_i = D(A_i,F_i)$ models all other factors changing the planktivore population, most importantly predation by adult piscivores. Since the planktivore population was initially small, we assumed $D$ to be linear in $F_i$.

We generated data to test this early warning signal with the previously published fishery model of Biggs \emph{et al.} \cite{Biggs_PNAS_2009}. Their model is a hybrid discrete-continuous system and explicitly models the adult piscivore, juvenile piscivore and planktivore populations. The intra-annual dynamics are continuous, modeling: the harvest of adult piscivores; the predation and control of planktivores and juvenile piscivores, respectively, by adult piscivores; predation on juvenile piscivores by planktivores; and the movement of both planktivores and juvenile piscivores between the foraging arenas and refugia. At the end of each year discrete update rules were applied that controlled the mortality of adult piscivores, maturation of the surviving juvenile piscivores into adult piscivores, and birth of more juvenile piscivores. There was an additive noise term on the planktivore population dynamics. For further details of the model including its mathematical formulation see Biggs \emph{et al.} \cite{Biggs_PNAS_2009}.

From the output of this simulation we recorded annual adult piscivore density, planktivore density, and piscivore catch. These observations were sufficient to calculate the Jacobian of Eqs.~\eqref{eq:Biggs},
\begin{equation*}
J = \left[
\begin{array}{cc}
1-m+r - (1-m)Q'^{(A)} - C'^{(A)} & -C'^{(F)} \\
-D'^{(A)} & 1 - D'^{(F)}
\end{array}
\right],
\end{equation*}
where we use the notation $\partial H/\partial x \equiv H'^{(x)}$. The eigenvalues $\lambda$ of the Jacobian, which may be complex numbers, are the solutions of $\det(J-\lambda I) = 0$ where $I$ is the ($2 \times 2$) identity matrix.

Our procedure was as follows. From the assumption of linearity in $A$, $Q'^{(A)}_i = Q_i/A_i$ can be calculated immediately. The terms $C_i$ and $D_i$ can be found by solving Eqs.~\eqref{eq:Biggs}. From the assumptions that $C$ is quadratic in $A_i$ and $D$ linear in $F_i$, it follows that $C'^{(A)}_i = 2C_i/A_i$ and $D'^{(F)}_i = D_i/F_i$. Finally, by solving linear Taylor expansions of the form
\begin{equation}
\label{eq:Taylor}
\Delta C_i = C'^{(A)}_i \Delta A_i + C'^{(F)}_i \Delta F_i,
\end{equation}
the derivatives $C'^{(F)}_i$ and $D'^{(A)}_i$ were estimated.

This generalized modeling approach required estimates of the adult piscivore mortality $m$ and reproduction rate $r$, and planktivore influx $R$ from refugia. We assumed the fishery manager could estimate $m = 0.4$ and $r = 1.1$ from knowledge of the fish species and the fishery. Effective values $m = 0.5$ and $r = 1$ can be derived from the actual simulation's equations \cite{Biggs_PNAS_2009}. The planktivore influx $R$ would be harder to estimate in practice. In the absence of such knowledge, the ecosystem manager could make use of the fact that the fish populations were initially stable. This knowledge constrains $R$ to the range $0.1$ to $1.5$, as values outside these ranges produce an eigenvalue with magnitude greater than one between times $t = 0$ to $10$. We used $R = 0.9$. Indeed, this knowledge could also be used to constrain $m$ and $r$ to the ranges 0.1 to 0.7 and 0.95 to 1.5, respectively. We found that any combination of $m$, $r$ and $R$ that gave an initially stable system robustly predicted the later critical transition.

Since the ecosystem approached the critical transition more slowly than the model with Allee effect above, to remove noise we found it necessary to smooth estimates of the gradients. For each new observation at time $t_i$, we fitted a second-order polynomial (since the method eventually results in differences of up to second order) to the record of each observed time series up to that time. Since it is the gradients at the current time that are most important, in the fit we applied a weight to each data point $j$ of $\exp[(t_j - t_i)/\tau]$ with $\tau = 10$. Gradients were then extracted directly from the coefficients of the polynomial fit. This simulated the calculation of early warning signals `on the fly', as new observations become available.

\section*{Tri-trophic food chain}
To obtain an early warning signal for this transition, we constructed a generalized model of the tri-trophic food chain
\begin{subequations}
\label{eq:trieq}
\begin{align}
\frac{\rm d}{dt}X_1 &= A(X_1) - G(X_1,X_2) \\
\frac{\rm d}{dt}X_2 &= G(X_1,X_2) - H(X_2,X_3) \\
\frac{\rm d}{dt}X_3 &= H(X_2,X_3) - M(X_3,\mu),
\end{align}
\end{subequations}
where $X_1$, $X_2$ and $X_3$ are the biomasses of the three species in the food chain, $A(X_1)$ is the primary production rate of $X_1$, $G(X_1,X_2)$ and $H(X_2,X_3)$ are the rates at which $X_2$ and $X_3$ consume $X_1$ and $X_2$, respectively, and $M(X_3,\mu)$ is the total mortality per unit time of the top-level predator $X_3$, with the argument $\mu$ denoting it is subject to a changing external influence.

We assumed: access to observations of all three biomasses $X_1$, $X_2$ and $X_3$, and the top-predator mortality $M(X_3,\mu)$; that this mortality is linear in $X_3$ so that $M = m(\mu) X_3$; and that the predation rates $G(X_1,X_2)$ and $H(X_2,X_3)$ are linear in their respective predator biomasses. These assumptions and observations were sufficient to calculate the Jacobian of Eqs. \eqref{eq:trieq}, and therefore the Jacobian's eigenvalues, by the same techniques as in the previous examples. The time series were first smoothed by the same filter as in the fishery simulation, but with $\tau = 20$. The approach could also easily be extended to the case of incomplete biomass conversion efficiencies, if estimates of those efficiencies were available.

\bibliography{GM,EWSCT}

\end{document}